\documentclass[10pt,letterpaper]{article}
\usepackage{opex3}
\usepackage{cite}

\begin{document}

\title{Ultrathin fiber-taper coupling with nitrogen vacancy centers in nanodiamonds at cryogenic temperatures}

\author{Masazumi Fujiwara$^{1,2,\dagger}$, Hong-Quan Zhao$^{1,2,\P}$, Tetsuya Noda$^{1,2}$, Kazuhiro Ikeda$^{3}$, Hitoshi Sumiya$^{3}$, and 
Shigeki Takeuchi$^{1,2,4, *}$}

\address{$^1$ Research Institute for Electronic Science, Hokkaido University, \\Sapporo, Hokkaido 001-0021, Japan
\\
$^2$ The Institute of Scientific and Industrial Research, Osaka University, \\Ibaraki, Osaka 567-0047, Japan
\\
$^3$ Advanced Materials R\&D Laboratories, Sumitomo Electric Industries, Ltd., \\Itami, Hyogo 664-0016, Japan
\\
$^43$ Department of Electronic Science and Engineering, Kyoto University, \\Kyoto 615-8246, Japan
\\
$\dagger$ Present address: School of Science and Technology, Kwansei Gakuin University, \\Sanda, Hyogo 669-1337, Japan
\\
$\P$ Present address: Chongqing Institute of Green and Intelligent Technology, \\Chinese Academy of Sciences, Chongqing 401120, China}
\email{takeuchi@kuee.kyoto-u.ac.jp} 



\begin{abstract}
We demonstrate cooling of ultrathin fiber tapers coupled with nitrogen vacancy (NV) centers in nanodiamonds to cryogenic temperatures.
Nanodiamonds containing multiple NV centers are deposited on the subwavelength 480-nm-diameter nanofiber region of fiber tapers.
The fiber tapers are successfully cooled to 9 K using 
our home-built mounting holder
and an optimized cooling speed.
The fluorescence from the nanodiamond NV centers is efficiently channeled into a single guided mode and
shows characteristic sharp zero-phonon lines of both neutral and negatively charged NV centers.
The present nanofiber/nanodiamond hybrid systems at cryogenic temperatures can be used as NV-based quantum information devices
and for highly sensitive nanoscale magnetometry in a cryogenic environment.
\end{abstract}

\ocis{(280.4788) Optical sensing and sensors; (240.6648) Surface dynamics; (230.3990) Micro-optical devices.} 


Quantum networks require different types of building blocks to realize their functions, such as information transfer,
processing, and storage \cite{kimble2008quantum}.
Photons are robust against decoherence and hence suitable for transmitting information between distant places
through optical fibers.
Solid-state quantum systems, such as quantum dots, defect centers in solids, and superconducting quantum circuits,
have large mutual interactions and are suitable for information processing, transducing, and storage \cite{xiang2013hybrid}.
Effective interaction between these quantum systems and photons in a fiber network is of paramount importance
to realize quantum networks.

Among the candidates, diamond nitrogen vacancy (NV) centers are very promising solid-state quantum systems for quantum networks \cite{doherty2013nitrogen}.
They exist in either bulk diamonds or nanodiamonds and show very stable emission without fluorescence blinking or bleaching.
The electronic states of NVs are spin triplet states and may be useful for quantum memory \cite{koshino2010deterministic}.
The electron spins show a long coherence time, which is important for coherent interaction between photons and NV centers \cite{bar-gill2013solid}.
In addition, at cryogenic temperatures, they have a lifetime-limited linewidth of optical transitions,
which is necessary for coherent excitation of NV centers \cite{tamarat2006stark, shen2008zero, chu2014coherent}.
For example, generation of quantum entanglement between photons and the electron spins
has been reported \cite{togan2010quantum}.
For the next step, it is critically important to improve the coupling efficiency between NV centers and photons passing through single-mode optical fibers.

Fiber tapers are promising nanophotonic devices for this efficient coupling between photons and quantum nanoemitters.
Their subwavelength diameters of 300--500 nm (so-called nanofiber) generate a significant evanescent field around the surface.
Through this evanescent field, quantum nanoemitters on the surface emit up to 20\% of the total fluorescence photons 
into a guided mode of the nanofibers \cite{fujiwara2011highly, yalla2012efficient, almokhtar2014numerical}.
Such NV/nanofiber hybrid devices have been reported in the context of single-photon sources \cite{schroder2012nanodiamond, liebermeister2014tapered} and magnetometers \cite{liu2013fiber}.
However, these studies were conducted only at room temperature and not at cryogenic temperatures.
Their cryo-cooling has been hindered by difficulties in cooling fragile ultrathin fiber tapers.

Here we report the first demonstration of cooling of NV/nanofiber hybrid devices to cryogenic temperatures.
We prepare nanodiamonds containing multiple NV centers and deposit them on a nanofiber region of fiber tapers with a diameter of 480 nm.
These fiber tapers are successfully cooled to 9 K using 
our home-built mounting holder
and an optimized cooling speed.
The fluorescence of the NV centers is efficiently coupled with the fiber tapers, showing
characteristic sharp zero-phonon lines (ZPLs) of both neutral (NV$^0$) and negatively charged NV (NV$^-$) centers.
The present NV/nanofiber systems at cryogenic temperatures can be used for NV-based quantum information devices such as quantum phase gates or quantum memories,
and for highly sensitive magnetometry in a cryogenic environment.


We synthesized high-quality type Ib single-crystal diamonds by using a temperature-gradient method under high-pressure and high-temperature
(HPHT) conditions of 5.5 GPa and 1350 $^\circ$C. 
Nitrogen concentration of the synthetic crystals was estimated to be about 100 ppm by FT-IR spectroscopy. 
The synthetic crystals were irradiated with a 4.6 MeV electron beam up to a dose of 100 kGy followed by an annealing at 800 $^\circ$C 
for 1 hour in vacuo. Then, the diamond crystals were crushed and separated in a mixture of acetone and ethanol.
The obtained nanodiamond powder was dispersed and suspended in ethanol and sonicated for an hour before use.
This nanodiamond suspension was spin-coated on thermally oxidized silicon substrates \cite{zhao2012suppression, zhao2012effect}.
The particle size distribution ranged from 500 nm to 2.0 $\mu$m according to scanning electron microscope (SEM) images.

\begin{figure}[t!]
\centering
	\includegraphics{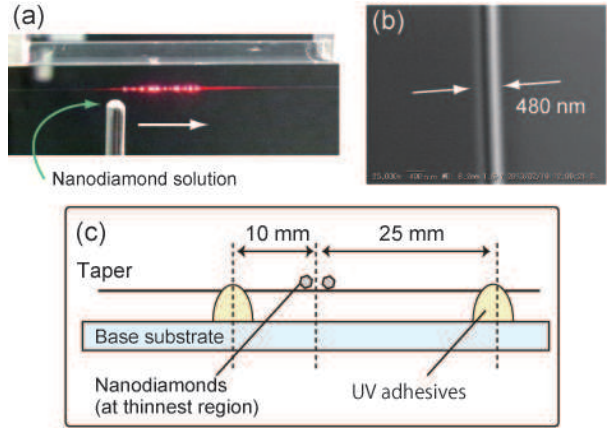}
	\caption{(a) Photograph of the dip-coating process. A glass rod having a droplet on top was moved in the direction indicated by the arrow to deposit the nanodiamonds.
				   (b) SEM image of a fiber taper with a diameter of 480 nm. (c) Schematic image of fiber taper mounted on base substrate.
				Note that the photograph in (a) shows symmetric mounting to explain the nanodiamond deposition procedure.
	}
	\label{fig2}
\end{figure}

%
%
The nanodiamonds show typical fluorescence spectrum of NV centers ranging from 550 to 750 nm with
sharp ZPLs at 577 and 639 nm at 4.2 K, which are ascribed to neutral (NV$^0$) and negatively charged NV (NV$^-$) centers, respectively.
We determined the content ratio of NV$^0$ and NV$^-$ in the nanodiamonds by measuring their fluorescence spectra  
using a home-made confocal fluorescence microscope equipped with liquid-helium cryostat 
(see \cite{zhao2012suppression, zhao2013observation} for the detail of the microscope).
The peak intensity ratio of NV$^0$ and NV$^-$ is $1.1 \pm 0.2$ for 17 nanodiamond particles.
Note that the content ratio does not affect the present experiment.


A small aliquot of the nanodiamond solution was placed on a facet of a small glass rod \cite{schroder2012nanodiamond}.
Fiber tapers were then dipped into the solution as shown in Fig.\ref{fig2}(a) (the fabrication of the fiber tapers is described below).
Nanodiamonds were deposited by moving the fiber taper along the fiber axis using a linear stage.
These nanodiamonds can be easily found by observing the scattering points of a red laser coupled in the fiber taper.
Note that taper  transmission is affected by the nanodiamond deposition.
The loss strongly depends on particle sizes, number of deposited particles, and particle shapes and is difficult to estimate.
Previous studies have reported a negligible loss for small nanoparticles ($\sim$ 10--30 nm) \cite{fujiwara2011highly, schroder2012nanodiamond} 
and 50 \% for a single 1.5-$\mu$m-sized particle \cite{gregor2009soft}.
In the present experiment, we use only the one side of the tapers due to the taper mounting method (see below)
and cannot measure the transmission loss by the dipcoating.
Nonetheless we successfully measure the large fluorescence signal through the taper compared with that through a high-NA objective at 9 K as described below.

Fiber tapers were fabricated from commercial single-mode optical fibers (Thorlabs, 630HP)
by a method described elsewhere \cite{fujiwara2011highly, fujiwara2011optical}.
Adiabatic tapering was confirmed, as the transmittance was larger than 0.9 during fabrication.
Figure \ref{fig2}(b) shows a SEM image of a 480-nm-diameter fiber taper.
The taper diameter was determined by an SEM.

One of the keys to successful cooling was the design of the holder for the nanofiber/NV center hybrid device.
The hybrid device floated on the surface of the base substrate (silica glass) supported by UV adhesives [Fig. \ref{fig2}(c)].
The contact positions between the UV adhesives and the fiber taper are critically important for successful cooling.
Ultrathin fiber tapers usually break abruptly, especially below the liquid nitrogen temperature (usually somewhere between 20 and 40 K),
whereas the optical transmission was conserved throughout the cooling process (when an optimized cooling speed was used, as described below).
We know empirically that the separation between the two UV adhesives is critical to avoid this sudden breaking of the fiber tapers.
Indeed, we have demonstrated cooling of 320-nm-diameter fiber tapers to 9 K by symmetrically narrowing this separation to 35 mm (each adhesive 17.5 mm from the center),
at the expense of optical loss \cite{fujiwara2012coupling}.
This time, we placed one UV adhesive 25 mm from the center and another at 10 mm from the center [see Fig. \ref{fig2}(c)].
This asymmetric mounting provides a high optical throughput for one end and effective narrowing of the separation,
thereby enabling successful cooling of ultrathin fiber tapers and conserving the high optical throughput.

\begin{figure}[t!]
\centering
	\includegraphics{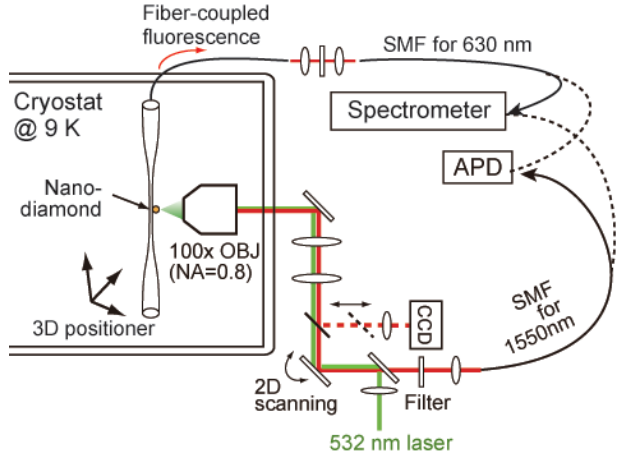}
	\caption{Schematic illustration of the experimental setup,
	including the cryostat and confocal microscope components.
	CCD: charge-coupled device, SMF: single-mode fiber, APD: avalanche photodiode.
	}
	\label{fig3}
\end{figure}

Figure \ref{fig3} shows a schematic of the setup of the fiber taper cooling experiment.
Fiber tapers having nanodiamonds on the surface were mounted
on a 3D axis translational stage (Attocube, ANPx51-LT and ANPz51-LT) inside a custom-made LHe flow cryostat \cite{fujiwara2012coupling}.
The sample chamber of the cryostat was filled with gas helium of half atmospheric pressure at room temperature,
which allows for quick thermalization of the fiber tapers.
A confocal microscope setup was built around the cryostat.
A microscope objective (Olympus, LMPLFLN 100x) with a numerical aperture (NA) of 0.8 was placed inside the cryostat.
A fiber-coupled continuous-wave 532 nm laser was used for excitation.
The polarization of the excitation laser was perpendicular to the taper axis, 
which is known to give minimal background fluorescence in the detection through the taper \cite{schroder2012nanodiamond}.
The taper was coarsely positioned using the 3D translational stage and fine-resolution raster scanning
by a 2D galvano mirror.
The fluorescence collected through the objective was coupled to a multimode fiber that acted as a pinhole (Thorlabs, 1550HP, core diameter $\sim$10 $\mu$m) and
was detected by either an avalanche photodiode (APD) or a spectrometer equipped with a charge-coupled device camera.
The fluorescence channeled into the fiber tapers was outcoupled once from the fiber end to be filtered out and incoupled again with single-mode fibers for detection.

The fiber tapers were cooled in the LHe flow cryostat.
Ultrathin fiber tapers easily lose transmission during the cooling process, and careful slow cooling is necessary.
The transmission is degraded by thermal expansion of the materials (e.g., nanofiber silica, UV adhesive, base mount silica).
The tapers were pre-cooled from room temperature to $\sim$80 K using liquid nitrogen at a cooling rate of $-$40 K h$^{-1}$.
The liquid nitrogen was then purged overnight, which eventually increased the temperature to 90 K.
They were then cooled to $\sim$20 K by liquid helium at a rate of $-$20 K h$^{-1}$, followed by a slower cooling to 
6.5 K at $-$6 K h$^{-1}$.
The temperature was then most stabilized at 9 K, where the temperature stability of our system was $\pm$0.05 K at 9 K, 
which was sufficient to prevent observable mechanical drift of the sample position.

\begin{figure}[t!]
\centering
	\includegraphics{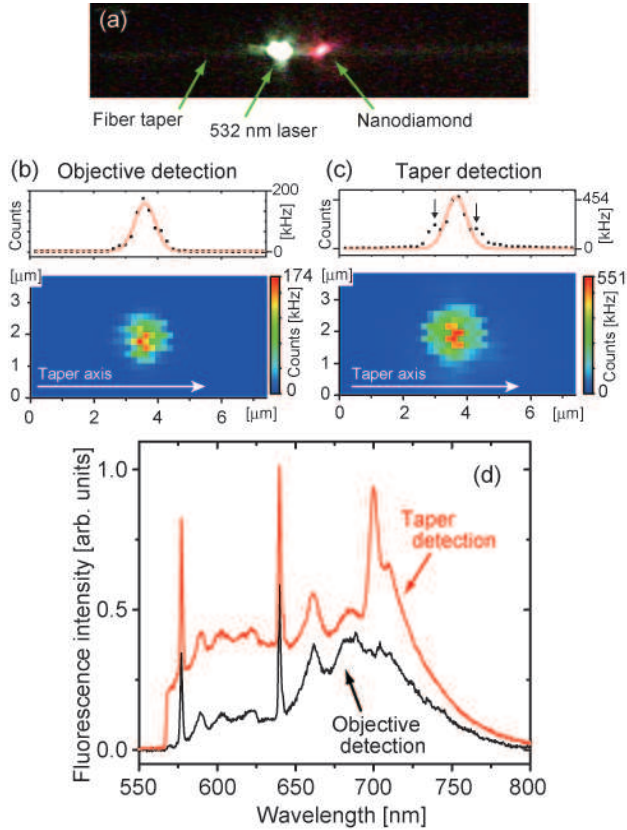}
	\caption{(a) Microscope photograph of the nanodiamond on the fiber taper at 9 K. A red laser is coupled in the fiber to
				visualize the nanodiamonds. Green scattering point is due to laser excitation.
				Raster scanning fluorescence images of this nanodiamond by detecting the fluorescence through (b) the objective and (c) the taper.
				(d) Fluorescence spectra of the nanodiamond for both detection methods.
	}
	\label{fig4}
\end{figure}
Figure \ref{fig4}(a) shows a microscope photograph of a 480-nm-diameter nanofiber with a deposited nanodiamond at 9 K.
A red laser was coupled to the taper to visualize the nanodiamonds.
The green scattering point at the center is the focus point of the laser.
After the taper was coarsely positioned, fine laser scanning was performed to obtain fluorescence raster scanning images.
Note that we have confirmed that there are no other red scattering points in the raster scanning area.

Figures \ref{fig4}(b) and (c) show fluorescence raster scanning images of the nanodiamond,
where the fluorescence was detected through the objective and the fiber taper, respectively.
The upper graphs of the scanning images are the cross sections of the images along the taper axis. 
A Gaussian fitting to the cross section in the objective detection gives FWHM of 0.76 $\mu$m. 
The cross section for the taper detection is shown in Fig. \ref{fig4}(c), where two side lobes (indicated by arrows in the figure) appear in the both sides of the central peak.
These side lobes may come from the distortion of the objective lens (described below) and is difficult to know the proper fitting function.
Rather fitting the data, we make Gaussian plots by fixing the linewidth to FWHM of the objective detection (0.76 $\mu$m) and 
the offset to the dark count signal of the far point from the peak.
The peaks are reproduced at the same lateral position in the both scanning images (Fig.  \ref{fig4}(b) and (c)).
We therefore think that we see the same nanodiamond in the both images of Fig. \ref{fig4}(b) and (c).
The detected photon count rate in the taper detection is 551 kHz [Fig. \ref{fig4}(c)], which is three times more than
that in the objective detection, 174 kHz [Fig. \ref{fig4}(b)].
This fact indicates the effectiveness of using fiber tapers for collecting fluorescence in cryogenic experiments.

Microscope objectives are normally designed for the usage at room temperature 
and needs a precise optical alignment inside the metallic body.
It is therefore inevitable to have a lower photon collection efficiency at cryogenic temperature because of 
the distortion coming from the large temperature difference of $\sim$ 300 K between room temperature 
and liquid helium temperature.
We indeed observed deformation of the point spread function of the laser excitation (reflected image of the laser beam)
as it cooled.
%
In contrast, fiber tapers do not show significant degradation of the optical throughput at our optimized cooling speed.
We have confirmed that a 340-nm-diameter fiber taper that has a UV adhesive separation of 50 mm (hence, high optical transmission) 
was able to be cooled to 20 K without degradation of the optical throughput (before suddenly breaking).
Although we were not able to monitor the optical throughput of the present one-sided fiber taper,
the optical throughput for one end is naturally conserved in this experiment.

Note that experimentally determining the coupling efficiency of the nanodiamond and the fiber taper is difficult at cryogenic temperature. 
Yalla et al reported a method \cite{yalla2012efficient} to determine the coupling efficiency by comparing the fluorescence intensity 
between the objective detection and the taper detection, in which some correction factors are necessary to know the optical throughput from the emitter to the detector via objective, mirrors, and so on.
It is however impossible to know the exact NA value or the transmission of the objective at 9 K, and only possible 
to conclude that the present taper detection more efficiently collects the fluorescence than the NA-0.8 microscope objective at 9 K. 

Figure \ref{fig4}(d) shows fluorescence spectra of the nanodiamond obtained using both detection methods.
Both of the spectra show strong NV characteristics.
Sharp ZPLs of NV$^0$ and NV$^-$ appear at 577 and 639 nm, respectively, for both detection methods.
All of these peaks have a linewidth of about 1.6 nm, which is comparable to the minimal wavelength resolution of the setup.
This indicates that the NV/nanofiber hybrid devices are indeed cooled to cryogenic temperature.
We compare the ratio of $NV^0 / NV^-$ between the two detection methods in order to further evaluate if the fluorescence comes from the same nanodiamond.
The  $NV^0 / NV^-$ ratio is 0.59 in the objective detection, whereas 0.82 in the taper detection. 
Since we excited the same nanodiamond, the relatively higher value in the taper detection indicates it may include some amount of background fluorescence.
The background fluorescence is especially prominent in the wavelength region of 560--620 nm in the taper detection, which explains the relatively higher value 
of $NV^0 / NV^-$ ratio in the taper detection.

The background fluorescence is observed only when we excite the nanodiamonds by a strong green laser.
It varies particle by particle depending on the size and shape, making it difficult to ascribe its origin to nanodiamonds or fiber tapers.
It may come from either of two or both. We therefore show the measured spectra without the background subtraction.
%
Note that there is a peak at around 700 nm only in the taper detection, which might be due to multimode interference \cite{fujiwara2011highly}.
However, these components would not affect most cryogenic experiments in quantum optics,
as only the sharp ZPLs are used to mimick atomic systems.
%

The present demonstration of coupling of ultrathin fiber tapers with multiple NV centers in nanodiamonds at cryogenic temperatures
can provide a testbed for NV-based fiber-integrated quantum information devices, such as quantum memories and quantum phase gates.
It is also expected to be useful for highly sensitive diamond magnetometry in a cryogenic environment.
The 480-nm-diameter nanofiber has a theoretical coupling efficiency for the two fiber ends of 20.0 \%, 13.4 \%,  and 12.8 \% 
for the radial, azimuthal, and axial dipole orientations, respectively (for an average of 15.4 \%) \cite{almokhtar2014numerical}.
These theoretical values have already been experimentally demonstrated using colloidal quantum dots \cite{yalla2012efficient}.
Our cooling technique may be extended to the smaller taper diameter of $\sim$ 300 nm where higher photon collection efficiency can be obtained 
for $\lambda$ = 637 nm of the ZPL of NV$^-$ centers \cite{almokhtar2014numerical}.
This coupling efficiency is further enhanced by integrating a fiber Bragg grating cavity either drilling the nanofibers or approaching photonic crystal cavities \cite{schell2015highly, nayak2011cavity, yalla2014cavity}.
The present cryogenic demonstration therefore offers a key technique to use these nanofiber-based devices for applications in quantum information and quantum metrology.


In summary, we demonstrated ultrathin fiber-taper coupling with multiple NV centers in nanodiamonds at cryogenic temperatures.
We prepared nanodiamonds containing multiple NV centers from Type Ib bulk diamond and deposited them on nanofibers with a diameter of 480 nm.
Having optimized the device structures and cooling speed, we successfully cooled the fiber tapers to 9 K.
The fluorescence of the NV centers was efficiently coupled with the fiber tapers, showing
characteristic sharp ZPLs of both neutral (NV$^0$) and negatively charged NV (NV$^-$) centers.
The present nanofiber/NV hybrid system at cryogenic temperatures can provide a testbed for NV-based fiber-integrated quantum information devices, such as quantum memories and quantum phase gates,
and magnetometry in a cryogenic environment.

\section*{Acknowledgements}
This research was funded in part by MEXT-KAKENHI Quantum Cybernetics (No. 21101007), JSPS-KAKENHI (Nos.
26220712, 23244079, 25620001, 23740228, 26706007, and 26610077), JST-CREST,
JSPS-FIRST, the Project for Developing Innovation Systems of MEXT, the G-COE
Program, and the Research Foundation for Opto-Science and Technology.

\end{document}